\documentstyle[12pt,psfig]{article}
\def\reference{\parskip 0pt\par\noindent\hangindent 0.5 truecm}

\begin{document}
%
%
\title{Spectral Ages of CSOs and CSS Sources }
%


\author{
Matteo Murgia
} 

\date{}
\maketitle

{\center Istituto di Radioastronomia del CNR, Via Gobetti 101,
I-40129
Bologna, Italy\\murgia@ira.cnr.it\\[3mm]

}

%
\begin{abstract}
This paper deals with the spectral ageing study of a representative
 sample of compact symmetric objects (CSOs) and compact steep spectrum (CSS) sources.
Observations reveal a distinctive high-frequency steepening of
the radio spectra of many of these sources. The existence of such a spectral feature
 is expected or may be naturally interpreted in terms of radiative ageing of
synchrotron emitting electrons. The small angular size of
CSS sources makes it relatively easy to measure their integrated spectra over a
 wide frequency range for a conspicuous number of objects.
For those sources whose emission is dominated by the mini-lobes, the integrated
 spectra can be used to constrain the source age. Assuming equipartition magnetic fields,
 the spectral ages we found are in the range from $10^{2}$ to $10^{5}$
 years. Multifrequency VLBA observations allow us to study the spectral
properties of two CSOs:  B1323+321 and B1943+546.
The case of B1943+546 is particularly interesting since for this source a kinematic
 age has been derived from the proper motion of the hot spots. We found that spectral
and kinematic ages agree within a factor of 2.
The overall results presented here confirm that the CSOs and CSS sources are indeed
 young objects.
Finally, we show some examples of compact sources characterised by
an
 extraordinary steep and curved spectrum. It is plausible that these
are relic sources in which the injection of fresh electrons has ceased for
 a significant fraction of their lifetime. These observations may indicate
either the presence of intermittent activity or a class of short-living
objects.

Throughout this paper we adopt ${\rm H_{\rm 0}}= 100\,h {\rm
\,km\, s^{-1} \,Mpc^{-1}}$ and ${\rm q_{\rm 0}}=0.5$.

\end{abstract}


\bigskip

%
%

\section{Introduction}
\label{Intro} This paper considers the spectral age of CSOs and CSS sources. The age is perhaps the most critical
parameter in understanding the nature of these intrinsically compact sources that have been alternatively
interpreted in terms of young radio sources ($t\simeq 10^3\sim 10^5$\,yr;
 Phillips \& Mutel 1982;
 Carvalho 1985; Fanti et al.\ 1995; Snellen et al.\ 1999) or
 ordinary age objects ($t\simeq 10^6\sim 10^7$\,yr) confined
 to subgalactic scales by a particularly dense interstellar medium (van Breugel, Miley, \& Heckman 1984).

Recently, two kinds of evidence are proving the youth scenario. First, the measurements of the hot spot separation
velocity in a dozen of the most compact CSOs (Owsianik \& Conway 1998; Polatidis \& Conway, this volume)
 demonstrate that the majority
 of these sources are certainly young. The second evidence constitutes the topic of
 this paper and is based on the derivation of the radiative age from the analysis of
the source spectrum.

The determination of the detailed evolution of the relativistic
 electron population responsible
 for the non-thermal radiation of radio sources is important because it allows one
 to acquire critical information concerning the physical conditions in the source
 emitting regions. In particular, if simple assumptions are satisfied, it is potentially
 possible to relate the curvature of the synchrotron spectrum to the age of the radiating
particle (Kardashev 1962; Kellermann 1964; Pacholczyk 1970; Jaffe
\& Perola 1974),
 a method known as `spectral ageing'.

Synchrotron losses  preferentially deplete high-energy electron populations, leading to
a steepening in the emission spectrum beyond a time and magnetic field dependent
 break frequency
\begin{equation}
\nu_{\rm br}\propto B^{-3}t_{\rm syn}^{-2}
\end{equation}
where $B$ and $t_{\rm syn}$ are the magnetic field strength and the radiative age, respectively.

Expressing the radiative age in years, the magnetic field
 in mG and the break frequency in GHz one finds\footnote{A number of assumptions enter in this formula. The magnetic field $B$ is
 considered to be constant.
 Synchrotron losses are supposed
 to dominate over expansion losses at the observing frequencies.
 The presumed magnetic field energy density in CSS sources and CSOs is of order magnitudes
 higher than the energy density of the cosmic microwave background photons; inverse
Compton losses
 are therefore neglected. Finally, it is assumed that
 the pitch angles $\theta$ between the electron velocity and the magnetic field
 direction are isotropically distributed and that the timescale for their continuous
re-isotropisation is much shorter than the radiative timescale,
$\langle (B \sin\theta)^{2} \rangle =\frac{2}{3} B^{2}$ (Jaffe \&
Perola 1974). }
\begin{equation}
\label{vbreak}
t_{\rm syn}=5.03\times 10^{4}\cdot B_{\rm mG}^{-1.5} [(1+z)\nu_{\rm br}]^{-0.5} \qquad \rm  (yr)
\end{equation}
where $z$ is the source redshift.

Although synchrotron losses always lead to a high-frequency steepening, the exact shape
of the spectral curvature depends on the evolution of the energy input into the
 relativistic electrons.
The `standard' synchrotron loss models consider the evolution of
the emission from an electron population with an initial power law
energy distribution. Two limiting situations can be considered: i)
a single injection of relativistic particle and ii) a continuous
particle injection. For both models the emission spectrum below
the break frequency,
 where radiative losses are unimportant, is a power law, $S_{\rm \nu} \propto \nu^{-\alpha_{\rm inj}}$. The models differ in the shape of their high-frequency steepening.
In the first case, energy losses lead to a sharp cut-off in the energy spectrum and, hence, to
 an exponential cut-off in the emission spectrum (JP model: Jaffe \& Perola 1974).
In the second case, the continuous energy supply limits the change in spectral index at
 the break frequency to $\alpha_{\rm h}=\alpha_{\rm inj}+0.5$ instead of an exponential drop
 (CI model: Kardashev 1962).
The JP model is adequate to describe the spectrum of the source if the regions where the electrons are
injected/accelerated can be distinguished from those where they age. Thus, if multifrequency
 images with proper resolution are available, the measure of the spectral variation
can be related to source dynamics.
The continuous injection model, instead, is supposed to describe the astrophysical
 situation in which these regions are not spatially resolved by the telescope beam.

\section{Integrated Spectra of CSS Sources}
\label{integrated} Because of their small angular size, CSS sources are unresolved for most scaled arrays, hence
it is very difficult to perform detailed spectral ageing studies
 across the mini-lobes of these objects. It is however possible to use sensitive interferometers
 such as the Very Large Array (VLA) and single-dish radio
 telescopes at very high frequencies ($> 10$\,GHz) to study their {\it integrated} spectra.
This permits us to perform a spectral ageing analysis for a significant number
 of sources and a large frequency range.
We have undertaken such a project with the study of two samples of CSS sources:
\begin{itemize}
\item[1)] the sample selected from the 3CR and the Peacock \& Wall (1981) catalogues
 (Fanti et al.\ 1995), 38 sources;
\item[2)] the B3-VLA CSS sample (Fanti et al.\ 2001), 87 sources.
\end{itemize}

\begin{figure}[t]
\centerline{
\psfig{figure=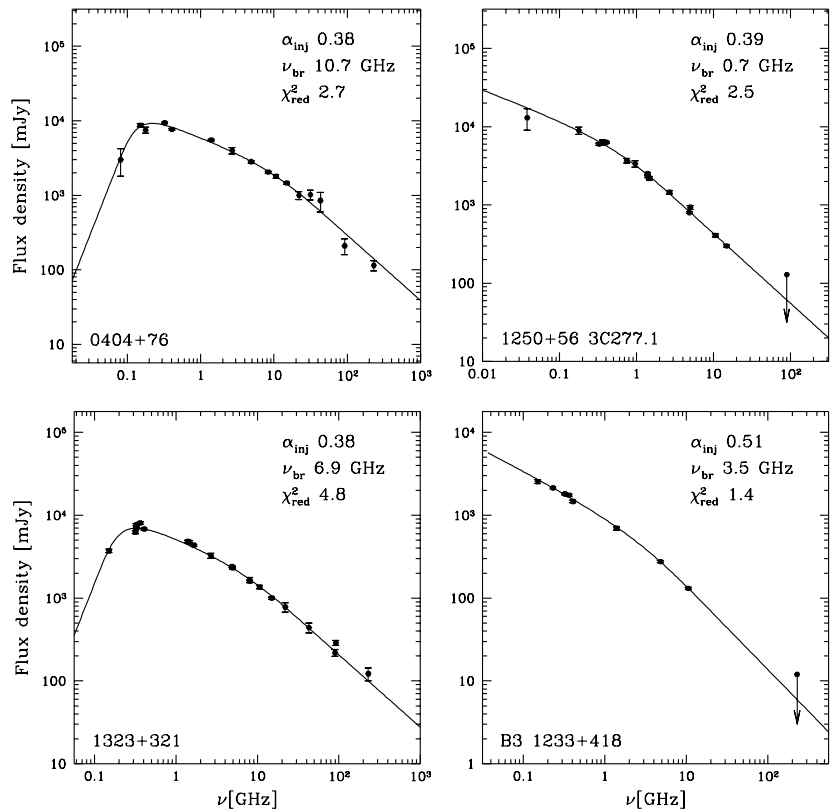,width=12cm}
}
\caption[]{Selected CSS integrated spectra. The solid line is the best fit of
 the CI model. The spectral features are of
the following types: a) a low frequency turnover (the most conspicuous deviation); b) a steepening at high
frequencies. The above deviations are interpreted as due to synchrotron self-absorption and particle energy
losses, respectively. } \label{fig1}
\end{figure}

We compiled flux densities at different frequencies from the
literature and from our own measurements (e.g. VLA 8.5 and
4.9\,GHz, Effelsberg 32\,GHz, Pico Veleta 230\,GHz). Examples of
source spectra are shown in Figure \ref{fig1}.

Most of the sources show significant deviation from the classical power law which describes a zero age transparent
synchrotron spectrum from a relativistic electron population with power law energy distribution. The deviations
from the power law can be summarised as follows: a) a low frequency turnover (the most conspicuous deviation), and
b) a steepening at high frequencies. High-frequency flattening, if any, is quite rare. The above deviations are
interpreted as due to synchrotron self-absorption and particle energy losses, respectively. Since we are dealing
with integrated spectra, we assume that the radio source evolution
 is described by a  continuous injection (CI)  model, where the sources are continuously
replenished by a constant flow of fresh relativistic particles with a power law energy
 distribution (see Section \ref{Intro}).

 By fitting a numerically computed CI spectrum to the data, one
obtains the break frequency $\nu_{\rm br}$, from which the source age is obtained
 if the magnetic field is provided (equation \ref{vbreak}). Since the spectral curvature of
 the CI spectrum is very gradual and occurs over more than two orders of magnitude
in frequency, the spectral measurements above 30\,GHz are of
fundamental importance to constrain the break frequency.

\begin{figure}[t]
\centerline{
\psfig{figure=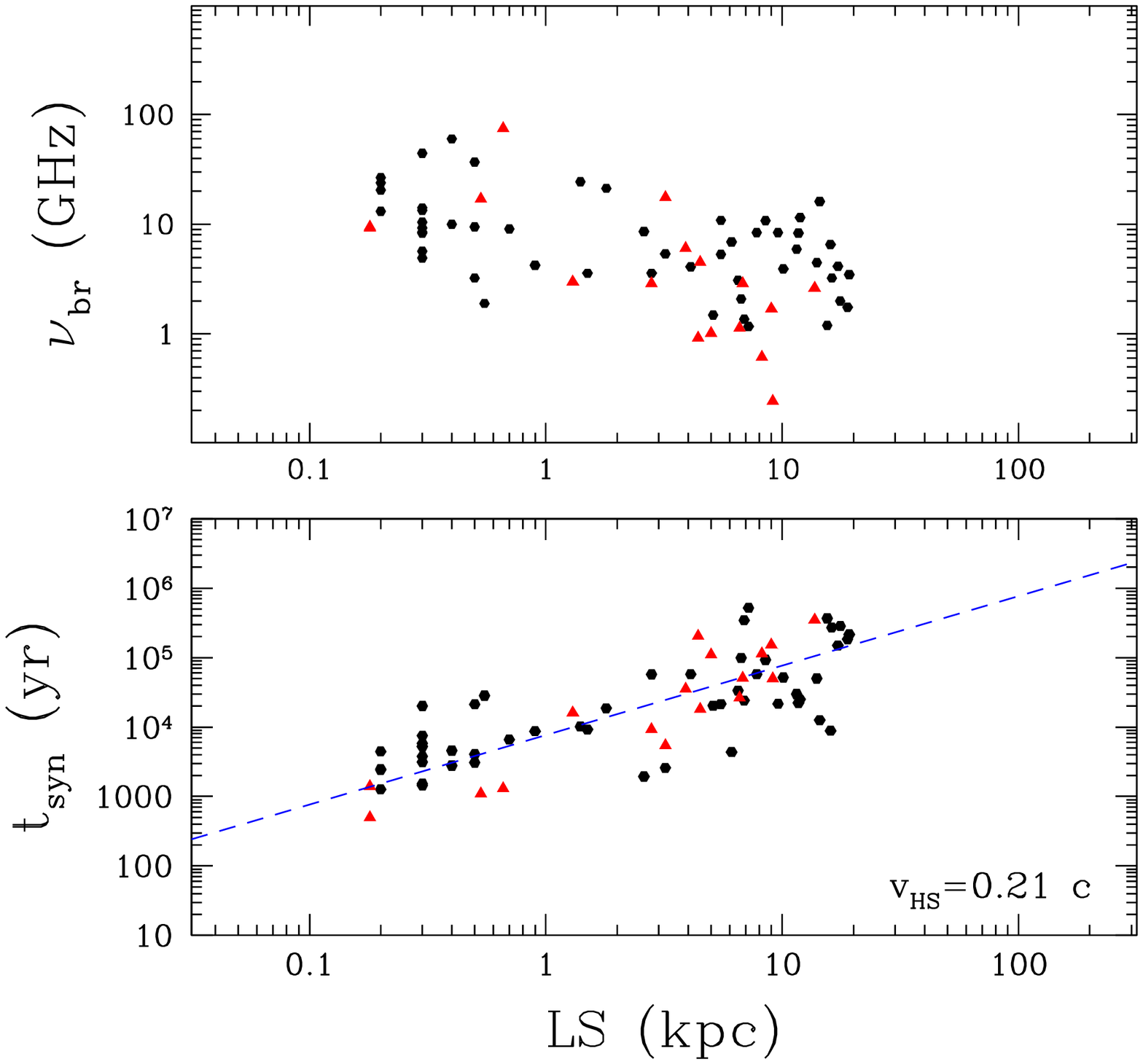,width=10.5cm}
}
\caption[]{Rest frame break frequency (top panel) and spectral age (bottom panel) as a
 function of the source linear size. Circles and triangles refer to the 3CR+WP and B3-VLA
 CSS samples, respectively. The diagonal dashed line correspond to a constant hot spot
advance speed of  $v_{\rm HS}=0.21 ~h^{-4/7}$ c.}
\label{fig2}
\end{figure}

One disadvantage of the integrated spectra is that the spectra of different
 source components are mixed up. If the source spectrum is dominated by a jet
or by hot spots, the radiative age likely represents the lifetime of the
 electrons in that component and it is expected to be less (perhaps much less) than
the source age.
In order to reduce this problem, we follow the same approach of Murgia et al. (1999),
 classifying the sources on the basis of their VLBI morphology.
In particular, we exclude all those sources from the analysis
 whose morphology is dominated by strong jets or hot spots.
In these cases, the radiative age derived from the integrated spectrum  might be
 completely unrelated to the source age. Only when the
lobes, which have accumulated the electrons produced over the
source lifetime, dominate the source spectrum, is the radiative
age $t_{\rm syn}$ likely to represent the age of the source.

The top panel of Figure \ref{fig2} shows the scaling of the rest
frame break frequency as a function of the source linear size
(LS). As expected, we found a good
 inverse correlation between $\nu_{\rm br}$ and LS: the larger a CSS source the smaller
 is the break frequency, i.e. the break frequency
behaves as an effective clock indicating  the source age. The
bottom panel of  Figure \ref{fig2} shows the spectral age,
calculated
 assuming  the equipartition magnetic field, as a function of the source linear size.
 The equipartition magnetic field has been computed with standard formulae (Pacholczyk 1970). We assumed particle energy equally distributed
into electrons and protons, a filling factor of unity, and ellipsoidal geometry, and we considered the radio
spectrum from 10\,MHz up to 100\,GHz. In these `lobe dominated' sources the estimated ages are in the range
between $10^{2}$ and $10^{5}~h^{-3/7}$ years and are well correlated with LS. The corresponding average hot spot
advance speed is $v_{\rm HS}= LS/2t_{\rm syn} =0.21 ~h^{-4/7}$ c. This value is relatively close, only about a
factor of 2 higher, to the expansion velocities of small size CSOs derived from VLBI observations. As a direct
consequence, $B_{\rm eq}$ seems a good estimate for the source magnetic field.

\section{Multifrequency Imaging of CSOs: the Cases of B1321+321 and B1943+546}
\label{multi} It is clear that multifrequency images
 are necessary in order to determine the spectral age of a single
 object with a better confidence level. Multifrequency imaging allows us
 to distinguish the source regions in which the electrons are injected from those
 in which electrons age, testing the dynamical models.

We present the preliminary results of a multifrequency analysis of the radio spectrum
 across the lobes of two bright CSOs: B1321+321 (Dallacasa et al. 1995) and B1943+546 (Polatidis et al. 1995).
Both sources are identified with galaxies at redshifts of $z=0.370$ and $z=0.263$,
respectively.
The sources have been selected on the basis of the following properties:
\begin{itemize}
\item[1)] VLBI morphology characterised by two, well resolved, microlobes with a
 linear angular size (LAS) $\sim 50$\,mas; the LAS of the proposed sources is such that we have a reasonable number of
resolution elements across the source, without loosing information on the more extended
features (lobes).

\item[2)] The integrated spectrum of each source is optically thin at frequencies
$>$500\,MHz and therefore any possible confusion between genuine
spectral steepening due to energy losses and the low frequency
turnover due to the absorption processes is excluded.

\item[3)] The integrated spectra show a clear high-frequency steepening and evidence for
 significant radiative losses.
\end{itemize}

\begin{figure}[t]
\centerline{ \psfig{figure=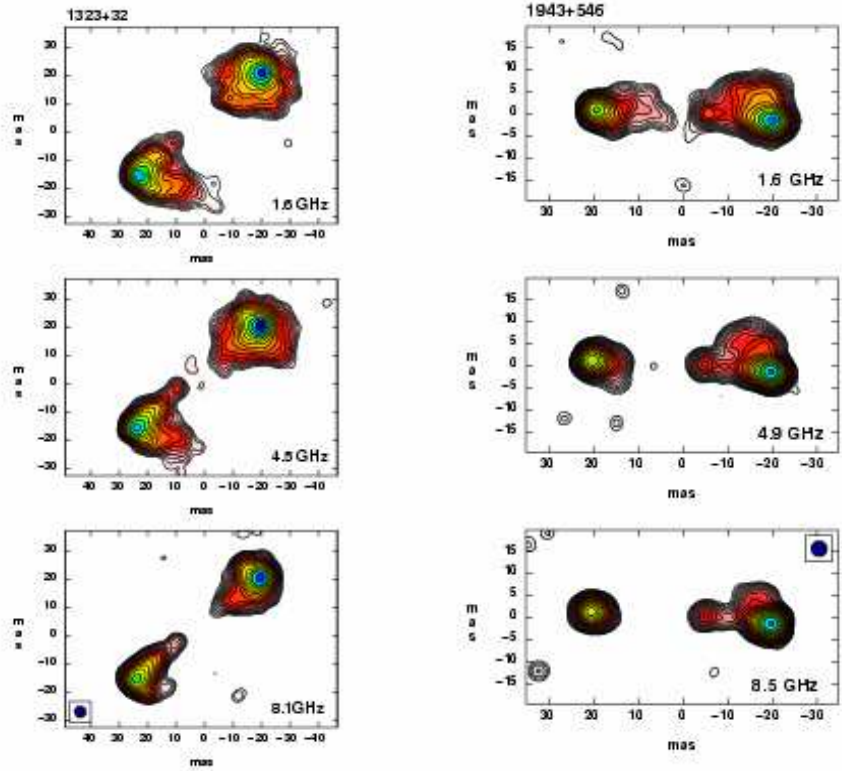,width=12cm} } \caption[]{Multifrequency images of B1321+321 (left panels)
and B1943+546 (right panels).} \label{fig3}
\end{figure}

We observed  B1321+321 and B1943+546 at 1.6, 4.5, 4.9, 8.1, and 8.5\,GHz with the VLBA+Effelsberg on 8 and 9 July
2000. We have produced images with reasonable {\it u-v} coverage at each frequency and such images are shown in
Figure \ref{fig3}. The final resolution is 4\,mas and 3.5\,mas for B1321+321 and B1943+546, respectively.

\subsection{B1321+321}
\begin{figure}[t]
\centerline{
\psfig{figure=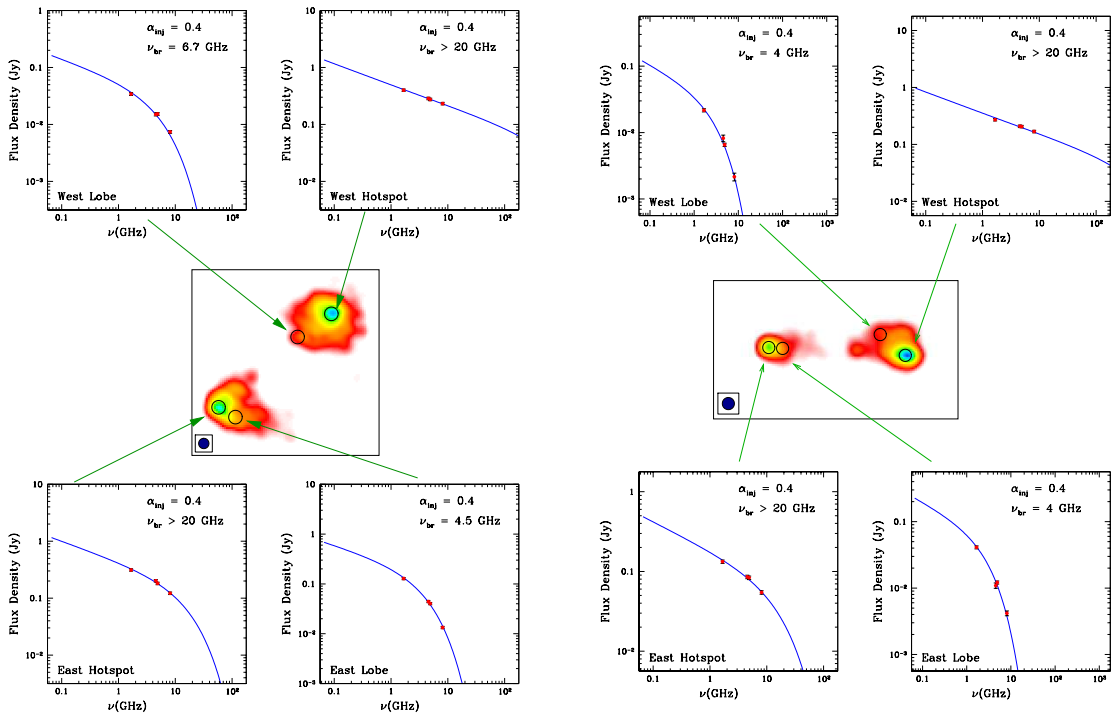,height=10cm}
}
\caption[]{Examples of spectral fits in the lobes of B1321+321 (left) and B1943+546 (right).}
\label{fig4}
\end{figure}

The morphology of B1323+321 is characterised by two symmetric microlobes and resembles the prototype of the
compact double radio source. Two bright hot spots are located at the edges of the lobes, as seen in the classical
extended and powerful sources. The SE lobe,
 which is closer to the core, has a tail emission to the SW
 with a shape typical of a back flow.
By fitting the JP synchrotron loss model (see Section \ref{Intro}) to the data we determined the variation of the
break frequency across the lobes of B1323+321. Examples of spectral fits are shown in Figure \ref{fig4} (left
panel). We found that both hot spots are characterised by a power law ($\nu_{\rm br} >
 20$\,GHz) spectrum with index $\alpha\simeq 0.4$. This value is consistent with the fit of
 the optically thin low frequency region of the integrated spectrum
(see bottom left panel of Figure \ref{fig2}), in agreement with
the idea that the injection of the young electrons occurs in these
regions. The radio spectrum in the NW lobe is clearly curved and
the break frequency decreases
 going from the hot spots to the core, a typical behaviour seen in many extended
 radio sources. The SE lobe spectrum steepens along the back flow tail but flattens in the direction
of the core, possibly due to the presence of the underlying jet. From the
 spectral fits in the NW lobe we found a minimum break frequency of $\nu_{\rm br} \simeq 7 $\,GHz. Assuming for the
lobe an ellipsoidal geometry of 30$\times$25\,mas, from standard equipartition formulae (Pacholczyk 1970) we found
$B_{\rm eq} = 4.3~{\rm mG}$. The corresponding synchrotron age is $t_{\rm syn} = 1860\,h^{-3/7}$ years. Given the
source size LS=180\,$h^{-1}$\,pc, the hot spot separation velocity, $v_{\rm sep} = 0.09$\,pc\,yr$^{-1}$ (i.e.
0.3\,$h^{-4/7}$c), results.

\subsection{B1943+546}
The observed structure of B1943+546 consists of a relatively weak core and two microlobes
 lying in the east--west direction, with the eastern lobe closer to the core. Both lobes contain a
 bright hot spot. The overall aspect of this CSO is virtually identical to that of a Fanaroff \& Riley (1974) type II source but scaled down by a factor of $\sim1000$. The spectral behaviour is similar to that
 of B1323+321 (see Figure \ref{fig4}, right panel). The injection spectral index in the hot spots
 is $\alpha\simeq 0.4$ and the break frequency decreases going from the lobe ends toward
the core. Again, this is consistent with the dynamical scenario in which the electrons at the centre of the source
are `older' than
 those closer to the hot spots, i.e. the source is expanding in time with the principal site of
particle acceleration being the hot spots. We restricted the spectral age calculation to the western lobe since it
is brighter then the eastern one and therefore it permits us to measure
 the break frequency much closer to the core.
The lowest break frequency we measured from the spectral fits in this lobe is
 $ \nu_{\rm br} \simeq 4 $\,GHz.
Assuming for the western lobe an ellipsoidal geometry of 20$\times$10\,mas we calculate an
 equipartition magnetic field strength of $ B_{\rm eq} = 6.7~{\rm\,mG}$.
The synchrotron age, $t_{\rm syn} = 1275\,h^{-3/7}$ years, results. Given the source size LS=107\,$h^{-1}$\,pc,
the hot spot separation velocity, $v_{\rm sep} = 0.086$\,pc\,yr$^{-1}$ (i.e. 0.28\,$h^{-4/7}$c), results.

This result agrees quite well with the observation of the hot spot proper motion in this CSO reported by Polatidis \& Conway (this volume), $v_{\rm sep} =0.26 ~h^{-1}$c.

\section{Is There Spectral Evidence for Intermittent Nuclear Activity
in CSS Sources?} \label{reborn} Several authors have suggested that the large fraction of CSS sources could be
explained in terms of recurrent activity in their nuclei (O'Dea \& Baum 1997; Reynolds \& Begelman 1997). Clues
supporting this view are given by the discovery of faint extended emission on kpc scales surrounding some CSOs
(Owsianik et al.\ 1998; Siemiginowska et al.\ 2002). We have seen that the continuous injection model generally
fits the integrated spectra of most CSS sources well. But what happens to the
 source spectrum if the injection of fresh electrons is {\it discontinuous}, as implied
 by the recurrent activity models?
We studied a discontinuous injection model in which active phases are
  cyclically followed by quiescent phases (Zappacosta 2000).
 The relative durations of these two periods
 are free parameters of the model. The model predicts the formation of
 a high-frequency exponential cut-off in the source spectrum just
after the beginning of each relic phase. This extreme break is quickly cancelled
 after the onset of the successive active phase and the ordinary CI
 spectral shape is recovered. Thus, in the integrated spectrum we have {\it two}
 break frequencies: $\nu_{\rm br}$, which keeps track of the overall source age, and
 a second higher break frequency, $\nu_{\rm br}^{\prime}$, which is indicative
 only of the actual quiescent phase duration.

\begin{figure}[t]
\centerline{
\psfig{figure=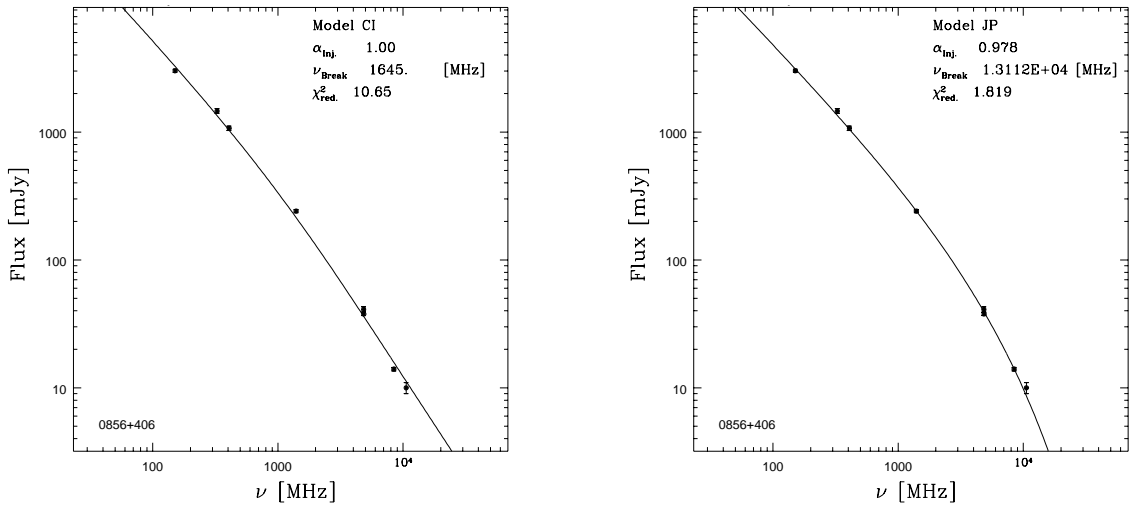,height=7cm}
}
\caption[]{The ultra-steep spectrum of the B3VLA-CSS 0856+406. The CI model (left) is
 not adequate to describe the extreme spectral curvature observed in this source.
 The spectrum is best fitted by the exponential cut-off of the Jaffe \& Perola (1974) model
(right). Also note the considerably high {\it injection} spectral index,
 $\alpha_{\rm inj} \simeq 1$, required by the fits.}
\label{fig5}
\end{figure}

We found in the B3-VLA CSS sample about 10\% of sources
characterised by
 integrated spectra whose steepening is too extreme for the CI model (see Figure \ref{fig5}).
The exponential cut-off observed in these spectra is incompatible with
 the moderate high-frequency steepening predicted by the CI model, suggesting that
the injection of fresh electrons in these sources may have ceased
for a significant fraction of their lifetime. In many cases, as
the example of B3 0856+406 shows (Figure \ref{fig5}), these
spectra have a very high {\it injection} spectral index
 $\alpha_{\rm inj} \geq 1$. A possible explanation for this is that we are
seeing a late relic phase, i.e. many cycles already passed and the first $\nu_{\rm br}$
 is now below our lowest observing frequency.

The discovery of such ultra-steep spectrum CSS sources has a double importance.
 First, it can be used to pinpoint relic CSS source candidates for
 further high resolution VLBI imaging. Second, an accurate determination
of the exact fraction of such steep spectra in CSS complete samples can in principle be used to constrain the
ratio between the active and relic timescales.

\section{Summary}
\begin{itemize}
\item[1.] From the study of integrated spectra we conclude that
 radiative ages of lobe dominated CSS sources are in the range of up to $10^{5}$ years, if
equipartition magnetic fields are assumed. The corresponding average hot spot advance speed is of the order of
0.2\,c.
\item[2.] Thanks to the increasing number of proper motion measurements
 in hot spots, CSOs will soon become important sources to test the synchrotron
ageing theory. We show that, in the  case of B1943+546, radiative and kinematic
 age are in very good agreement.
\item[3.] The very steep and curved spectra of some B3-VLA CSS sources suggest that the injection
 of fresh electrons in these sources has ceased for a significant fraction of
 their lifetime.
\end{itemize}

%
%





\section*{Acknowledgments}


I would like to thank my collaborators Carlo Stanghellini, Daniele Dallacasa,
 Carla \& Roberto Fanti, and Karl-Heinz Mack for their continuing support.

\section*{References}






\reference Carvalho, J.C.\ 1985, MNRAS, 215, 463

\reference Dallacasa, D., Fanti, C., Fanti, R., Schilizzi, R.T., Spencer, R.E.\ 1995, A\&A, 295, 27

\reference Fanaroff, B.L., \& Riley, J.M.\ 1974, MNRAS, 167, 31

\reference Fanti, C., Fanti, R., Dallacasa, D., Schilizzi, R.T., Spencer, R.E., Stanghellini, C.\ 1995, A\&A, 302, 317

\reference Fanti, C., Pozzi, F., Dallacasa, D., Fanti, R., Gregorini, L., Stanghellini, C., Vigotti, M.\ 2001, A\&A, 369, 380

\reference Jaffe, W.J., \& Perola, G.C. 1974, A\&A, 26, 423

\reference Kardashev, N.S.\ 1962, SvA, 6, 317

\reference Kellermann, K.I.\ 1964, ApJ, 140, 969

\reference Murgia, M., Fanti, C., Fanti. R.,  Gregorini, L., Klein, U., Mack, K.-H., Vigotti, M.\ 1999, A\&A, 345, 769

\reference O'Dea, C.P., \& Baum, S.A.\ 1997, ApJ, 113, 148

\reference Owsianik, I., \& Conway, J.E.\ 1998, A\&A, 337, 69

\reference Owsianik, I., Conway, J.E., \& Polatidis, A.G.\ 1998,
A\&A, 336, 37L

\reference Pacholczyk, A.G.\ 1970, Radio Astrophysics (San Francisco: Freeman \& Co.)

\reference Peacock, J.A., \& Wall, J.V.\ 1981, MNRAS, 194, 19

\reference Phillips, T.J., \& Mutel, R.L.\ 1982, A\&A, 106, 21

\reference Polatidis, A.G., Wilkinson, P.N., Xu, W., Readhead, A.C.S., Pearson, T.J., Taylor, G.B., Vermeulen, R.C.\ 1995, ApJS, 98, 1

\reference Polatidis, A.G.,\& Conway, J.E.\ 2002, PASA, 20, in press

\reference Reynolds, C., \& Begelman, M.\ 1997, ApJ, 487, L135

\reference Snellen, I.A.G., Schilizzi, R.T., Miley, G.K., Bremer, M.N., Röttgering, H.J.A., van Langevelde, H.J.\ 1999, NewAR, 43, 675

\reference Siemiginowska, A., Bechtold, J., Aldcroft, T.L., Elvis, M., Harris, D. E., Dobrzycki, A.\ 2002, ApJ, 570, 543

\reference van Breugel, W.J.M., Miley, G.K., \& Heckman, T.A.\ 1984, AJ, 89, 5

\reference Zappacosta, L.\ 2000,  Diploma Thesis, University of Bologna

\end{document}